\documentclass[twocolumn,pra,aps,showpacs]{revtex4}

\usepackage{psfrag,graphicx}

\usepackage{dcolumn}
\usepackage{amsmath,amssymb}
\usepackage{bm}
\usepackage{pstricks}

\def\half{\textstyle\frac{1}{2}}

\begin{document}



\title[Short Title]{ Critical Lines and Massive Phases
in Quantum Spin Ladders with Dimerization}

\author{
J. Almeida$^{\star}$, M.A. Martin-Delgado$^{\star}$ and G.
Sierra$^{\ast}$
 }
\affiliation{ $^{\star}$Departamento de F\'{\i}sica Te\'orica I,
Universidad Complutense. 28040 Madrid, Spain.
\\
$^{\ast}$Instituto de F\'{\i}sica Te\'orica, C.S.I.C.- U.A.M.,
Madrid, Spain. }

\begin{abstract}
We determine the existence of critical lines in
dimerized quantum spin ladders in their phase diagram
of coupling constants using the finite-size DMRG algorithm.
We consider both staggered and columnar dimerization patterns,
and antiferromagnetic and ferromagnetic inter-leg couplings.
The existence of critical phases depends on the precise combination
of these patterns. The nature of the massive phases separating the
critical lines are characterized with generalized string order parameters
that determine their valence bond solid (VBS) content.
\end{abstract}


\maketitle


\section{Introduction}
\label{sect_intro}

The issue of the existence of massive or critical phases in quantum spin systems
has motivated a great deal of study in strongly correlated systems since the
seminal work on the Haldane conjecture \cite{haldane82}. This issue became a
central problem when more complicated arrays of spins chains, known as quantum
spin ladders, were discovered experimentally in cuprate materials
that exhibit high-$T_c$ superconductivity when they are appropriately doped
\cite{dagotto_rice96}, \cite{strongly}.

From a more fundamental viewpoint, the study of critical and
massive phases in quantum spin ladders offer the possibility of a
testing ground for studying complicated quantum many-body effects,
that in some instances underly the physics of unconventional
phases of matter \cite{dagotto}. A variety of spin ladders with
different number of legs have been synthesized based on cuprate
materials, like $SrCu_2O_3$, $Sr_2Cu_3O_5$ etc. \cite{hiroi_et_al91}, \cite{azuma_et_al94}
or in other family of compounds $La_{4+4n}Cu_{8+2n}O_{14+8n}$ \cite{batlogg_et_al95},
and they present typically antiferromagnetic
rung couplings among the legs of the spin ladders. It is also possible
to find materials with less conventional ferromagnetic rung couplings as well
like in certain chemical compounds like PNNNO and PIMNO \cite{ferro_experiment}.
In addition, the new tools to study strongly correlated systems based on optical
lattices open the possibility of implementing a variety of quantum spin systems
including ladders \cite{greiner}, \cite{OLs}.

The phase diagram of quantum spin ladders with staggered
dimerization was conjectured on the basis of analytical
non-perturbative methods \cite{snake_ladders96} like the
non-linear sigma model (NLSM) complemented with additional
information from the weak and strong coupling limits in the rung
coupling constants and dimerization parameters. Later, a series of
different approximate  analytical studies \cite{kotov_et_al99},
\cite{wang_nersesyan00} have been favorable for the existence of a
critical line in the simplest case of a 2-leg spin ladder with
staggered dimerization. Also, some preliminary numerical methods with
the Lanczos algorithm \cite{lanczos_2leg98}, \cite{okamoto03} have shown support for this fact
for ladders with small size.
Despite these several studies, a complete non-perturbative numerical analysis of these staggered
low dimensional quantum spin systems
have remained as an open problem.

In this paper we study a 3-leg quantum spin ladder with different
types of rung couplings (either antiferromagnetic or
ferromagnetic) and different dimerization patterns. These ladders
are complex enough so as to serve as paradigmatic examples for
testing the conjectured phase diagrams \cite{snake_ladders96}.
In order to achieve a conclusive answer to the conjectured phase diagrams
for this system, we resort to a non-perturbative numerical tool like the
DMRG method \cite{white92}, \cite{white93}, \cite{hallberg06}, \cite{scholl05},
\cite{dmrg_book} in particular we employ its finite-size version based on the sweeping
procedure to improve the convergence of the iterative steps. The characterization
of the different phases separated by the critical lines is performed with a
DMRG calculation of generalized string order parameters (SOP) \cite{oshikawa92} that were introduced
to distinguish between massive phases in dimerized spin chains \cite{affleck_haldane87}
even when the localized spins in the chain take on half-integer values. These SOPs are
extensions of the originally non-local vacuum-expectation values introduced for integer spins
and the like \cite{sop1}, \cite{sop2}.

The elucidation of the existence of a critical line in the phase diagram
of quantum spin ladders with staggered dimerization is not straightforward
when the ladders have end points as in the open boundary conditions geometry
demanded by the standard DMRG method. Thus, we have to resort to numerical analysis
of the low-lying spectrum of excitations in order to extract the correct
gap in the bulk of the system when studying the universal properties of these
ladders in the thermodynamic limit (length going to infinity) \cite{FAF2legladder}.

The string order parameter is a theoretical construct in condensed matter
that allows the characterization of massive phases with a VBS structure.
It has never been measured experimentally. However, with the engineering
of optical lattices it would be possible to address a direct measurement
of this important quantity \cite{OLs}.

This paper is organized as follows:
 in Sect.\ref{sect_model} we introduce the two patterns of dimerization
 in 3-leg ladders, one being columnar \eqref{model1} with ferromagnetic rung
 couplings, while the other is alternating \eqref{model2} with antiferromagnetic
 rung couplings. In the former, we establish a helpful connection with the
 $S=\frac{3}{2}$ antiferromagnetic alternating chain in a certain strong coupling
 limit. We recall several conjetures about the phase diagram of both 3-leg Heisenberg
 models \cite{snake_ladders96} that motivate their study with DMRG in order to
 clarify them.
 In Sect.\ref{sect_critical} we perform the DMRG calculations of the low-energy gaps
 in both models of 3-leg ladders as a function of the various coupling constants.
 With this information we can establish the phase diagram and thus we establish
 the validity of the conjectured diagram \cite{snake_ladders96}. In addition, we can
 give a precise location of the critical lines and we find qualitative differences
 between these critical lines in each model.
 In Sect.\ref{sect_phases} we introduce generalized string order parameter to characterize
 the nature of the massive phases separated by the critical lines found in the previous
 section. These string orders are measured with DMRG techniques and we show that they are
 a valuable tool for detecting VBS states in dimerized quantum spin ladders with different
 patterns of dimerization.
Sect.\ref{sect_conclusions} is devoted to conclusions.
In appendix \label{sect_appendix} we study with DMRG the $S=\frac{3}{2}$ alternating spin ladder
and its generalized string order parameters. This case appears as a limiting case in the study
of the ferromagnetic 3-leg ladder with columnar dimerization and it is used as a guiding example
to find the phase diagram.

\section{Quantum Hamiltonians for 3-Leg Spin Ladders with Dimerization}
\label{sect_model}

One of the main interests established in \cite{snake_ladders96} was the existence
of an interplay between 3-leg ladders with a columnar dimerization pattern and ferromagnetic
inter-leg, or rung, couplings on one side, and 3-leg ladders with alternating dimerization
and antiferromagnetic rung couplings, on the other. The point was that both arrangements of
3-leg ladders should exhibit critical lines, while neither their precise location was known
nor the nature of the massive phases they separated was determined. This is the open problem
that we address here by means of the DMRG method. To this end, we start introducing both arrays
of ladders since they are the candidates to exhibit critical lines.

\begin{figure}[h]
\includegraphics[scale=1.0]{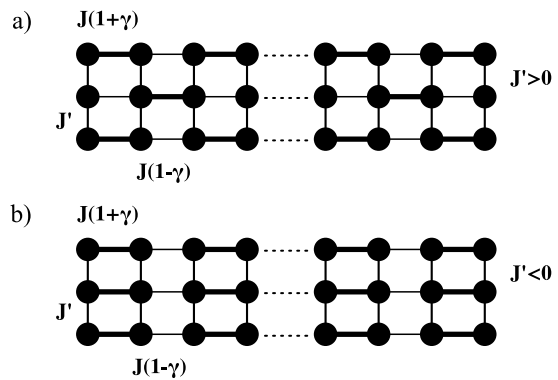}
\caption
{
Pictorial representation of the Hamiltonians corresponding \textit{a)} to the completely
antiferromagnetic ($J'>0$) model \eqref{model2} with alternated staggering and \textit{b)} 
the ferromagnetic ($J'<0$) model \eqref{model1} with columnar staggering.
}
\label{ladderpics}
\end{figure}

Thus, we are mainly interested in studying a possible connection between two different
arrangements of spins $\half$ forming a 3-leg ladder which interact via Heisenberg terms.
 Both of them combine bonds of different
strength parameterized by a constant $\gamma$ and two different types of coupling constants, $J$
for the Heisenberg interaction between spins along the legs,
and $J'$ for similar interactions between the rungs of the legs.
Since the physics of the problem depends only on the ratio $J'/J$, the constant $J$
will from now on be given a fixed antiferromagnetic value $J=1$. The other constant
$J'$ will have positive sign in one of the models and negative in the other one. In addition to this
difference, both models will also differ in the staggering pattern as we shall discuss below.

We next describe the model corresponding to the region with $J'<0$, with the bond alternation
pattern such that every one of the three legs begins with a strong bond followed
by a weaker one, that is, the bond alternation follows a {\em columnar pattern}. The explicit
Hamiltonian of this arrangement is
\begin{equation}
\begin{split}
H_{\textrm{F}}= &J\sum_{\ell=1,2,3} \sum_{i=1}^{L-1}(1-(-1)^i\gamma)\mathbf{S}_{i}(\ell) \cdot
\mathbf{S}_{i+1}(\ell) \\
&+J'\sum_{i=1}^L \mathbf{S}_{i}(1) \cdot
\mathbf{S}_{i}(2) +
J'\sum_{i=1}^L \mathbf{S}_{i}(2) \cdot
\mathbf{S}_{i}(3),
\end{split}
\label{model1}
\end{equation}
where $L$ denotes the longitudinal length of the 3-leg ladder, $\mathbf{S}_{i}(\ell)$ is
a spin-$\half$ operator located at the site $i$ of the $\ell$-th leg with $0<i\le L$,
$\gamma$ is the dimerization parameter that sets the relative strength of the bonds that
interact with Heisenberg coupling constant $J$ along the leg of the ladder and $J'$ on
the rungs of the legs.

Since the coupling among legs is ferromagnetic, we know that for values
 $\vert J'\vert\gg1$ this model converges to an effective $S=3/2$ staggered spin chain.
 Studies with the  NL$\sigma$M method predicts \cite{affleck88}
 for this chain the existence of three critical points in the interval $\gamma\in[-1,1]$
placed at $\gamma_c=\pm 2/3$ and $\gamma_c=0$. Numerical studies showed that in
fact these points correspond to $\gamma=\pm 0.42$ and $\gamma=0$ \cite{takahashi96}, \cite{yamamoto97}.
 Since the $S=3/2$ dimerized spin chain  gives us valuable information about the expected massive
phases of the ladder in the strong ferromagnetic regime, and since there exist some subtleties regarding
the string order parameters that may lead to confusion, we have included a numerical study
 of this chain in appendix \ref{sect_appendix}.

For the completely antiferromagnetic regime $J'>0$, we will use another staggering pattern
that differs with respect to the previous one. In this case only the first and third leg begin with a strong bond,
while the second one begins with a weak one, that is, the bond alternation is not columnar
anymore but still follows a regular pattern that is called {\em alternating}. The Hamiltonian of this model is
\begin{equation}
\begin{split}
H_{\textrm{AF}}= &J\sum_{\ell=1,2,3} \sum_{i=1}^{L-1}(1+(-1)^{i+\ell}\gamma)\mathbf{S}_{i}(\ell) \cdot
\mathbf{S}_{i+1}(\ell) \\
&+J'\sum_{i=1}^L \mathbf{S}_{i}(1) \cdot
\mathbf{S}_{i}(2) +
J'\sum_{i=1}^L \mathbf{S}_{i}(2) \cdot
\mathbf{S}_{i}(3),
\end{split}
\label{model2}
\end{equation}
with the same conventions as before. A pictorial representation of this Hamiltonian together 
with the Hamiltonian of the ferromagnetic model is shown in fig. \ref{ladderpics}.

With this arrangement, it is also  possible to effectively and approximately map the model
onto a NL$\sigma$M and again this formalism
predicts a critical behaviour in the phase diagram of the couplings
$J'/J$ vs. $\gamma$ \cite{snake_ladders96}. However, this behaviour is only
reliable in the strong coupling limit $J'/J \gg 1$.s
 More specifically, it predicts a critical
curve running from the point $(\gamma=2/3,J'=0)$ to $(\gamma=1, J'=4/5)$ and another one which
is the mirror reflection of the latter with respect to the $J'$ axis. These predictions however
 shall be considered only as qualitative approximations of the real behaviour of the system.
 In this particular case, it is evident that the critical line must cut the
 $\gamma$ axis exactly at $\gamma=0$ since that point corresponds to two decoupled $S=1/2$ staggered Heisenberg chains.
 However, this behaviour is missed by the NL$\sigma$M technique.
 On the other hand, this model has not apparent limits which can give us a hint on the phases
that give raise. In section \ref{sect_critical} however, our DMRG computations will give strong evidence of
their nature.

Despite the differences in both models, i.e. different sign of $J'$ and different staggering
pattern, there are various features that connect them. First of all and more important is that
at least in the line $\gamma=0$ both models are constinuosly related as we vary $J'<0$ to
 $J'>0$. On the other hand, according to NL$\sigma$M, the first model is critical only in the
region $J'>0$ while the second one has only critical lines in the complementary part $J'<0$.
 This dual-like behaviour combined with the expectation that the ground state of both models
is a valence bond solid, rouse the belief  about the possibility of
establishing a connection among their phase diagrams \cite{snake_ladders96}.
We shall see to what extend these expectations are fulfilled with the help of the DMRG technique
and the generalized string order parameters.

\section{Ground state degeneracy and existence of critical lines}
\label{sect_critical}

Massive quantum phases are characterized by an energy gap from the ground state (degenerate or
not) to the first excited state. On the contrary critical phases are characterized
 by a gapless spectrum between these energy levels. We have used the finite-size DMRG
 algorithm to compute the low energy levels and thus the corresponding gaps in order to
 identify the gap in the bulk of the system when we send the length of the 3-leg ladders to
 infinity (thermodynamic limit).

It is known that in the case of integer spin chains, some configurations of VBS states can break a hidden
$Z_2\times Z_2$ symmetry \cite{oshikawa92} that makes the ground state degenerate.
This degeneration is a reflection of the spin-end effects in a antiferromagnetic Heisenberg chain
of $S=1$ spins \cite{Kennedy90}.
In systems other than the parent Hamiltonians of the VBS states, but close enough
 to this picture and when using open boundary conditions, typically this degeneracy is
approximate and the energy of the near-degenerate states decays with the size exponentially
to a unique infinite volume ground state.

\begin{figure}[h!]
\includegraphics[scale=1.0]{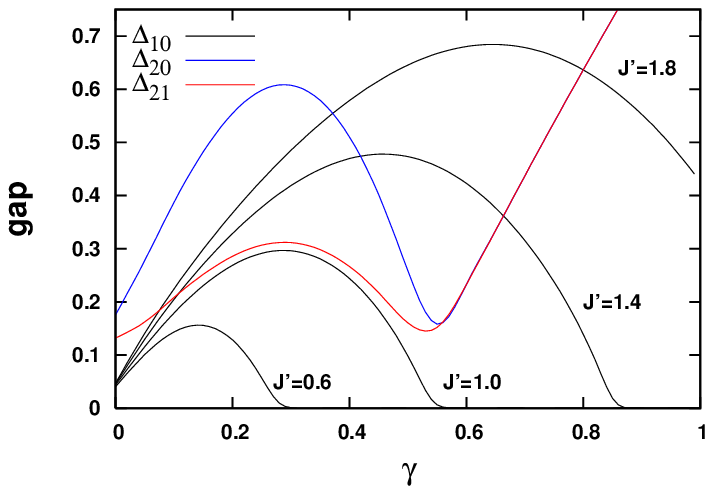}
\includegraphics[scale=1.0]{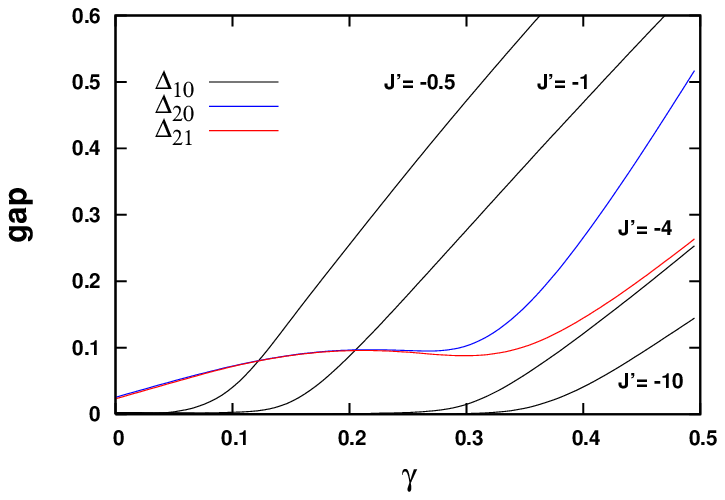}
\caption
{
(Color online)Energy gaps between the ground state and two
 first excited states in model \eqref{model2} 
(\textit{Up}) and model \eqref{model1} (\textit{Down}). For the sake of clarity 
$\Delta_{20}$ and $\Delta_{21}$ are shown only for one value of $J'$. Namely $J'=1.0$ 
(\textit{Up}) and $J'=-0.4$ (\textit{Down}). It can be seen how the gap $\Delta_{21}$
 indeed represents properly the gap of the spectrum irrespective of the degeneration of the
ground state.
}
\label{gap_xy}
\end{figure}

Fig. \ref{gap_xy} shows the energy differences between the ground state and two first excited
states in both models \eqref{model1} and \eqref{model2}. We observe in this
figure that $\Delta_{10}$
accounts for the degeneracy mentioned in the previous paragraph: in the completely 
antiferromagnetic model \eqref{model2} the
first excited state is clearly above the ground state in the phase corresponding to low values
of $\gamma$, while it is degenerate in the rest of the $\gamma$ interval.
 On the contrary,
 the ground state of the ferromagnetic model \eqref{model1} is degenerate for low 
values of $\gamma$ while it
has a finite gap in the rest.
As for $\Delta_{20}$, it corresponds in fact to the gap of the spectrum in the degenerate
regions while  it clearly differs from the gap $\Delta_{10}$ in the non degenerate
regions. Finally, the energy difference $\Delta_{21}$ coincides
 in both regions and both models with the gap of the massive phases.
 This holds true up to slight deviations due to finite
size effects and irrespective of the degeneracy of the ground state,
Thus, we may conclude that this gap $\Delta_{21}$ is in fact the same one that survives
when using periodic boundary conditions.

As mentioned in the previous section, the existence of critical lines (characterized
 by a gapless spectrum) in both
models \eqref{model1} and \eqref{model2} is supported by arguments coming from
 the strong coupling limit in the case of model \eqref{model1}, and NL$\sigma$M
 valid in both of them. These arguments are however not conclusive. In this section
we will prove numerically that these lines exist and we give an accurate estimation of its shape
and location.
To this end, we will find the critical values $\gamma_c(J')$ that make the gap
 $\Delta_{21}(\gamma_c(J'))$ vanish. It is very important at this point to
emphasize that an exactly vanishing value of the gap shall only be
attained in the thermodynamic limit. For finite size systems, the magnitude of the
gap remains finite and gets closer and closer to zero as we increase the size.
 In our case however, critical points are separating
massive gapped phases and therefore, for large enough sizes but still computationally
feasible, the gap at these points
attain a local minimum value and can be accurately computed.
\begin{figure}
\includegraphics[scale=1.0]{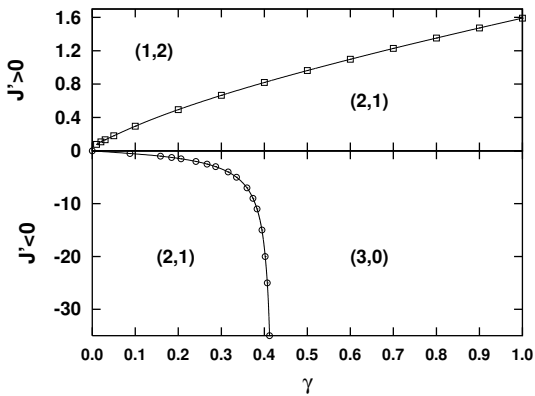}
\caption
{
Critical curves of model \eqref{model1} with $J'>0$ and alternated staggering, and
model \eqref{model2} with $J'<0$ and columnar staggering. Each point in the critical lines
has been obtained keeping one parameter fixed and finding the value of the other parameter 
that minimizes the gap $\Delta_{21}$. Computations have been performed on ladders of size
$L=3\times150$ retaining $m=400$($J'>0$ region) and $m=450$($J'<0$ region) states of the 
density matrix. 
}
\label{critical_lines}
\end{figure}

Fig.\ref{critical_lines} shows the critical region computed for both models
\eqref{model2} and \eqref{model1}. In order to compute this curves, we have
used the finite DMRG algorithm in ladders of $L=3\times 150$ sites. For the
completely antiferromagnetic model \eqref{model2} we
retained $m=400$ states of the density matrix and a Lanczos tolerance equal to $10^{-9}$.
The ferromagnetic model \eqref{model1} turned out to be numerically more demmanding and
we set $m=450$ and the tolerance equal to $10^{-10}$. Two sweeps of DMRG were enough in both models
to make the energies converge. These results clearly confirm the conjectured phase diagrams for these
3-leg models \cite{snake_ladders96}.

The solid lines in fig. \ref{critical_lines} are only a guide for the eye. We have however used
our numerical data to stimate the best fit to that critical lines. For the region $J'>0$ 
we have used a simple potential function of the form 

\begin{equation}
J'_c=a\gamma_c^r
\end{equation}

The best value of each parameter has been obtained performing a least square fit and 
are equal to $a=1.59\pm0.01$ and $r=0.72\pm0.01$.
As for the critical line in the semiplane $J'<0$, we have used for the region close to
 the vertical asymptota a relation of the form

\begin{equation}
J'_c=\frac{C}{(\gamma_c-a)^s}
\end{equation}

And the values that best fit the data have been found to be $C=0.38\pm0.02$, $s=1.07\pm0.02$ 
and $a=0.427\pm0.001$. Notice that the value of this last parameter is in good agreement with 
previous computations of the critical point of the $S=3/2$ alternating dimerized chain.

\section{Massive phases and generalized string order parameters}
\label{sect_phases}
\begin{figure}[h]
\includegraphics[scale=1.0]{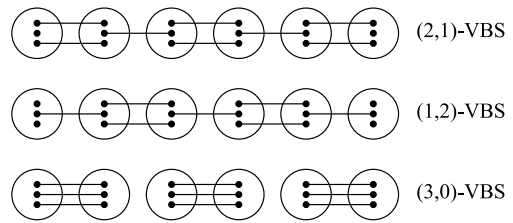}
\caption
{
Valence bond solid diagrams of the phases that give raise the models discussed in this paper.
Each small solid circle and line represents both spin-1/2 variable and a singlet pair 
repectively. The large open circles represent the symmetrization of the spin 1/2-variables on
each leg to create a spin-3/2 variable. 
}
\label{3legVBS}
\end{figure}

In this section we will characterize the quantum phases that
appear in the phase diagram in Fig.\ref{critical_lines}. To
achieve this goal, we will resort to the generalized string order
parameter \cite{oshikawa92}, \cite{sop2}, which are able to detect
the VBS state character of dimerized spin systems even when the
local spins take on half-integer values. These
parameters are generalizations for arbitrary complex phase of the
original string order \cite{sop1}, \cite{sop2} parameter first
 proposed  for the case of integer spin $S=1$.
Resorting to the VBS picture, massive phases corresponding to
valence bond solids
 can be denoted according to the number of valence bonds formed with the contiguous sites, i.e,
 one particular valence bond solid can be denoted as $(m,n)$-VBS with
 $m+n=2S$.

 For instance, we have already mentioned in the previous section that in the strong
 coupling limit, the columnar dimerized 3-leg ladder \eqref{model1} effectively becomes
 a $S=\frac{3}{2}$ alternating spin chain. Thus, in this case we have $m+n=3$.

The definition of the generalized string order parameter extended
to our particular three leg ladder with arbitrary size $L=3\times\ell$ is
\begin{equation}
O_{\textrm{str}}(\theta)=\Big \vert\lim_{j-i\rightarrow\infty} \langle
S^z_{2i}\textrm{exp}(i\theta\sum_{k=2i}^{2j-1}S_k^z)S_{2j}^z\rangle\Big\vert
\end{equation}
with $0<i<j<\ell/2$ and $S^z_i=S^z_i(1)+S^z_i(2)+S^z_i(3)$. It
is actually not necessary to consider $j-i\gg1$ to obtain
accurate values of the parameter and typically
 a value of $i-j$ of some few tens is enough to give values very close to the infinite
limit value, considered that $i$ and $j$ are well within the bulk and
far away from the edges. Thereby,
 for convenience we will
work with the parameter defined as
\begin{equation}
O_{\textrm{str}}(2i,2j,\ell,\theta)=\Big\vert \langle
S^z_{2i}\textrm{exp}(i\theta\sum_{k=2i}^{2j-1}S_k^z)S_{2j}^z\rangle\Big\vert
\label{SOPdef}
\end{equation}
with $i$,$j$ and $S^z_i$ defined as before.

 It has been shown \cite{oshikawa92} that the generalized
 string order parameter
evaluated in $\theta=\pi$ acts as an order parameter since it vanishes
or not depending on the number of bonds $n$ being odd or even.
Moreover, the shape of the string order parameter in the region
$\theta\in[0,2\pi]$ provides us with valuable information about the VBS character
of the phases  since the number
 of zeros in this range coincides with the number of bonds $m$ \cite{oshikawa92}.

\begin{figure}
\includegraphics[scale=1.0]{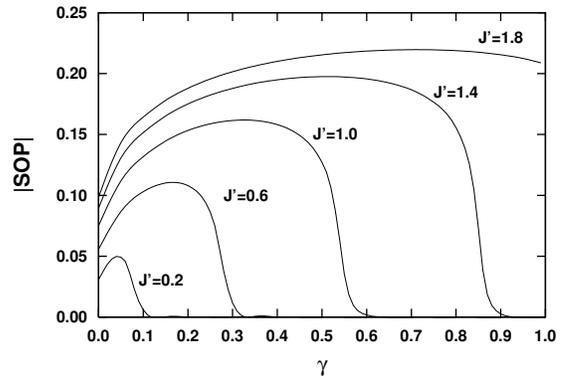}
\caption { Generalized string order parameter
$O_\textrm{str}(i=20,j=42,\ell=80,\theta=\pi)$ computed in the completely antiferromagnetic
model \eqref{model2} with alternated staggering for several fixed values of $J'$.  Regions where
the string order parameter vanishes correspond to a different
quantum phase from that where
 it is non null. Notice that there exists a certain $J'$ above which the system only exhibit
 one quantum phase irrespective of the value of $\gamma$. This value corresponds to the
$J'$ coordinate with $\gamma_c=1$ of the critical line of figure
\ref{critical_lines}. } \label{SOPvsGamma_aferro}
\end{figure}

\begin{figure}
\includegraphics[scale=1.0]{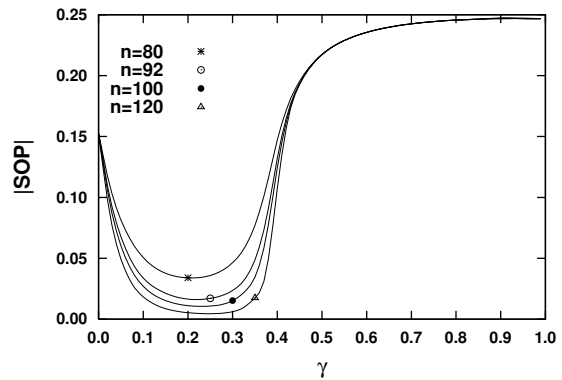}
\caption { Generalized string order parameter
$O_\textrm{str}(i=\ell/2-24,j=\ell/2+26,\ell=n,\theta=\pi)$ computed in the
strong ferromagnetic regime $J'=-25$ of the ferromagnetic model \eqref{model1} with
columnar staggering. See the text for explanations. }
\label{SOPvsGamma_ferro}
\end{figure}

Fig. \ref{SOPvsGamma_aferro} shows the parameter $O_\textrm{str}$
computed in the completely antiferromagnetic model \eqref{model2}
in the whole range of $\gamma$ for various values of $J'$. The
operator clearly distinguishes regions where it is finite from
others where it vanishes. Moreover for a fixed value of $J'$, the
value of $\gamma$ where it decays
to zero coincides \cite{comment1}
with the critical value
$\gamma_c$ corresponding to that value of $J'$ in the critical
line of Fig. \ref{critical_lines}. On the other hand, Fig. \ref{SOPvsGamma_ferro}
shows $O_\textrm{str}$ computed in the
ferromagnetic model \eqref{model1} in the strong coupling regime.
For strong values of the ferromagnetic coupling $J'$ we should
expect that our ladder behave like an effective
 $S=3/2$ alternating spin chain. Indeed, we can observe that the string order parameter is clearly non-vanishing
above $\gamma=0.42$, the critical point of the chain. As for
the region below this point,  the tendency of the string order
parameter is to decay to zero as we increase the size of the
system, except for the point $\gamma=0$ and its vicinity. In fact,
this behaviour is anomalous since $\gamma=0$ is critical in the
chain and therefore the string order parameter should vanish. In
the appendix we have addressed this issue with the pure $S=3/2$
alternating dimerized chain. Our study in the chain explains the behaviour of the ladder
and shows that indeed the string order parameter decays to zero
also at this point. The decay rate is however slower and it is not
enough to increase the size of the ladder. Fig. \ref{S32_SOP} (down)
shows that we have additionaly to consider sites $i$ and $j$ further
and further apart to make the string order parameter decay at
$\gamma=0$.

\begin{figure}
\includegraphics[scale=1.0]{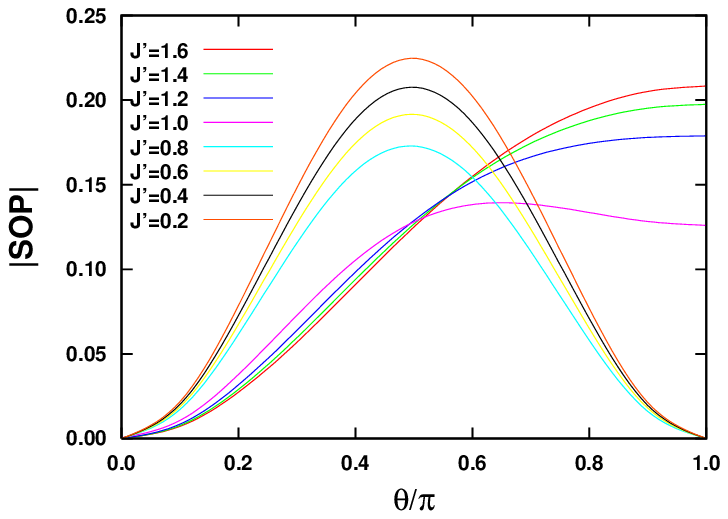}
\includegraphics[scale=1.0]{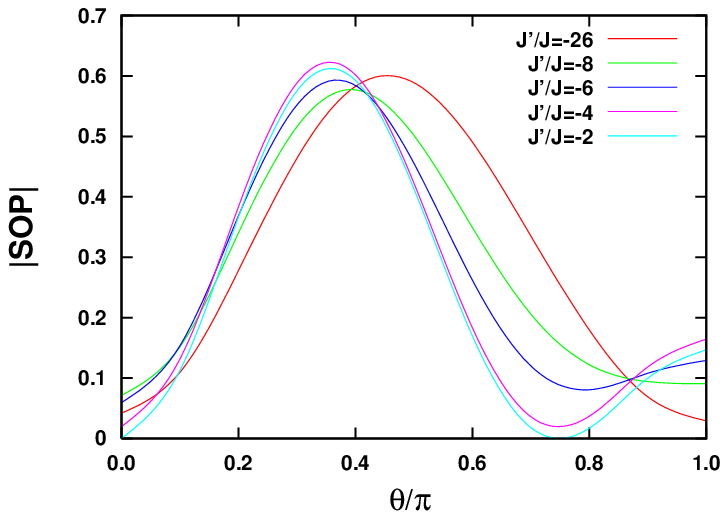}
\caption {(Color online)Generalized string order
parameter $O_\textrm{str}(i=20,j=42,\ell=80,\theta)$
 computed in the completely antiferromagnetic model
at $\gamma=0.5$ (\textit{Up}) and the ferromagnetic model at
$\gamma=0.35$ (\textit{Down}). According to
fig. \ref{critical_lines} this values cut the critical
lines at
 $(\gamma_c=0.5, J'_c=0.96)$ and $(\gamma_c=0.35, J'_c=-6.1)$ respectively. It can be observed
 that both graphs show a change in the number of zeros near these points.
}
\label{SOPtheta_ferro}
\end{figure}

With the considerations explained in the previous paragraph, we have
proved that the string order parameter \eqref{SOPdef} evaluated in
$\theta=\pi$ is valid to identify quantum phase transitions in
both models \eqref{model1} and \eqref{model2}. Now we want to
extend the study to the whole $\theta$ domain to test the nature
of the
 massive phases. In fact, from definition \eqref{SOPdef}
 $O_\textrm{str}(i,j,\ell,\theta)=O_\textrm{str}(i,j,\ell,\theta+\pi)$
and hence we can restrict the study to the range
$\theta\in[0,\pi]$. We will study first the ferromagnetic model
\eqref{model1} with columnar staggering. As we commented in the previous sections, we can
guess the phases of the diagram going to the strong coupling limit
$\vert J'\vert\gg1$. In this regime the ferromagnetic coupling
among rungs is the leading interaction and the ladder transforms
into an effective $S=3/2$ alternating spin chain. Resorting
 to continuity arguments, the phases of the ladder must be the same that appears in the strong
coupling limit, i.e, a $(2,1)$-VBS in the region
$0<\gamma<\gamma_c$ and a $(3,0)$-VBS when $\gamma_c<\gamma\le 1$.
In fig.\ref{SOPtheta_ferro} (down) we show $O_{\textrm{str}}$
computed in the $\theta$ domain. All the curves appearing in the
figure correspond to a fixed $\gamma=0.35$. According to
 Fig. \ref{critical_lines} this value of $\gamma$ cuts the critical line in a certain value
of $J'$ and so we should notice a qualitative change in the curves
in Fig. \ref{SOPtheta_ferro}. This change can be observed since
plots corresponding to very negative $J'$ have a local minimum at
$\theta=\pi$, while it changes to become a maximum as we move towards
$J'$ close to zero. Hence, the number of zeroes in the domain
$\theta\in[0,2\pi)$ moves from one to two. In fact,
 the string order parameter is not strictly equal zero in our graphs, but this fact has
 been already pointed out in spin chains \cite{yamamoto97} and conclusively proved that it
was due to finite size effects. The values closer to zero are
attained considering larger sizes.
 According then to the VBS notation and our DMRG results,  we can label
 the phases of model \eqref{model1} as $(2,1)$-VBS in the region $0<\gamma<\gamma_c$ and
$(3,0)$-VBS for $\gamma_c<\gamma\le1$ as expected from the
knowledge of the $S=3/2$ chain.

The completely antiferromagnetic model is more difficult to guess
a priori the quantum phases that gives raise to or even if its
ground states are valence bond solids. We have used again the
generalized string order parameters to check the nature of the
phases. Fig.\ref{SOPtheta_ferro} (up) shows the string order
parameter as a function of $\theta$ and a fixed value $\gamma=0.5$
which cuts the critical line. It can be seen that
 indeed the SOP behaves as expected for a valence
bond solid VBS state and two phases can be identified attending to the
number of zeros in this domain. For higher values of $J'$ the SOP
has a maximum at $\theta=\pi$ and only one zero in the region
$0\le\theta<2\pi$. As we consider lower values of $J'$, more
precisely in the interval from $J'=1.0$ to $J'=0.8$, the SOP at
$\theta=\pi$ falls abruptly to zero and therefore the SOP has two
vanishing values in the aforementioned interval. From these results we
can conclude firstly, that both massive phases can be properly
described as valence bond solids and also they can  be identified
as a $(1,2)$-VBS  for $0\le\gamma<\gamma_c$ and a $(2,1)$-VBS for $\gamma_c<\gamma\le 1$.

\section{Conclusions}
\label{sect_conclusions}

We have given a precise meaning to the conjectured phase diagrams \cite{snake_ladders96}
corresponding to 3-leg Heisenberg ladders with columnar dimerization and ferromagnetic
rung couplins on one side, and similarly for alternating dimerization with antiferromagnetic
couplings among the rungs. Although both models exhibit critical lines, their qualitative form
is different: in the former, the critical line approaches an asymptota at the critical value
$\gamma_c=0.42$ were it effectively becomes an alternating $S=\frac{3}{2}$ spin ladder.
On the contrary, the latter model does not exhibit any asymptota but the critical line meets
the wall $\gamma=1$ of the phase diagram.

Moreover, we have also clarified the valence-bond-solid nature of the massive
 phases separated by the critical
lines in the phase diagram. In this regard, we have found that the generalized string order
parameters are a very good tool for characterizing VBS state phases in ladders with a variety
of dimerization patterns. Our results are based on extensive calculations using the finite-size
DMRG technique.

\bigskip
\noindent {\em Acknowledgements}:
Part of the computations of this work were performed with the High
Capacity Computational Cluster for Physics of UCM (HC3PHYS UCM), funded
in part by UCM and in part with FEDER funds.
We acknowledge financial support from  DGS grants  under contracts BFM 2003-05316-C02-01,
FIS2006-04885, and the ESF Science Programme INSTANS 2005-2010.

\appendix


\section{The $S=3/2$ staggered spin chain}
\label{sect_appendix}
It was already pointed out by Oshikawa \cite{oshikawa92} that half-integer systems are amenable
to have many order parameters. We want to compare in this section two order parameters which
 despite differing slightly in their definition have in fact quite different behaviour and may 
even lead to confusion. We will see that both parameters capture the quantum phase transition
 of the chain, but only one can go further and give evidences of the quantum phases themselves.

\begin{figure}
\includegraphics[scale=1.0]{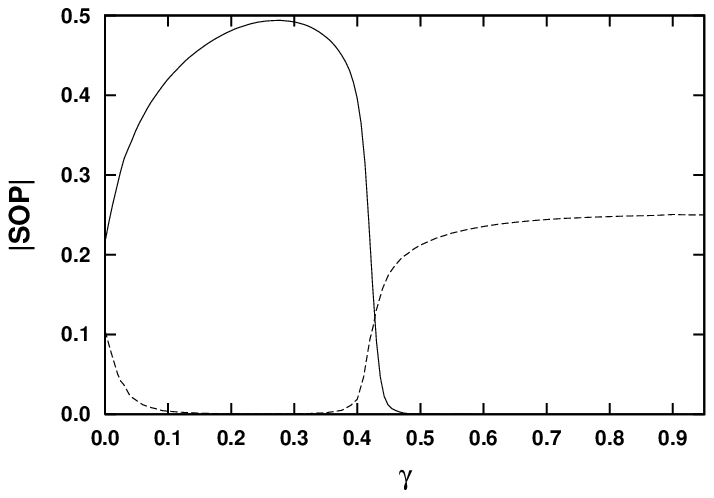}
\includegraphics[scale=1.0]{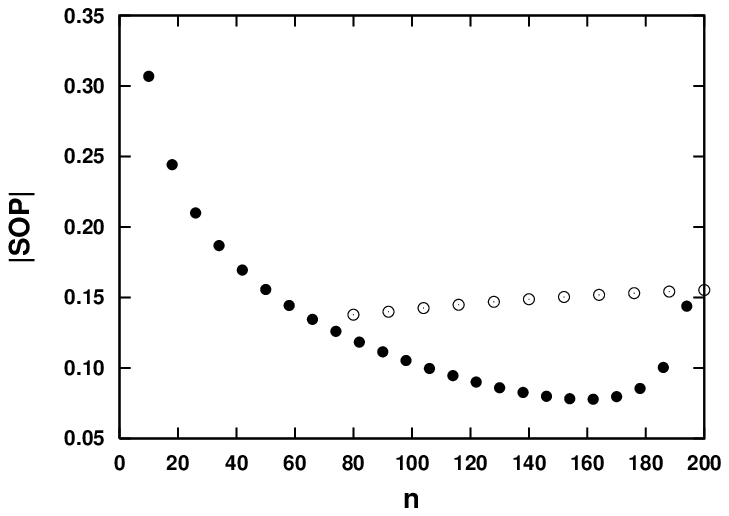}
\caption
{
\textit{Up}: string order parameters $O^\textrm{odd}_\textrm{str}(i=51,j=131,L=180,\theta=\pi)$
(solid line)
and $O^\textrm{even}_\textrm{str}(i=52,j=130,L=180,\theta=\pi)$(dashed line)
 computed in a $S=3/2$ alternating dimerized chain.
\textit{Down}: string order parameter $O^\textrm{even}_\textrm{str}$ at $\gamma=0$ varying
the total length of the chain $O^\textrm{even}_\textrm{str}(i=n/2-24,j=n/2+26,L=n,\theta=\pi)$
 (empty circles) and varying the distance
 $i-j$, $O^\textrm{even}_\textrm{str}(i=(L-n+1)/2+1,j=i+n-1,L=200,\theta=\pi)$ (solid circles).  
}
\label{S32_SOP}
\end{figure}

The string order parameters are defined as

\begin{equation}
O^\textrm{even}_\textrm{str}(\theta,L)=
\Big\vert
\lim_{j-i\rightarrow\infty}
\langle S_{2i}^z\textrm{exp}(i\theta\sum_{\ell=2i}^{2j-1}S_\ell^z)S_{2j}^z\rangle
\Big\vert
\end{equation}

with $0<i<j\le L/2$, and

\begin{equation}
O^\textrm{odd}_\textrm{str}(\theta,L)=
\Big\vert
\lim_{j-i\rightarrow\infty}
\langle S_{2i+1}^z\textrm{exp}(i\theta\sum_{\ell=2i+1}^{2j}S_\ell^z)S_{2j+1}^z\rangle
\Big\vert
\end{equation}

with $0\le i<j< L/2$.

\begin{figure}
\includegraphics[scale=1.0]{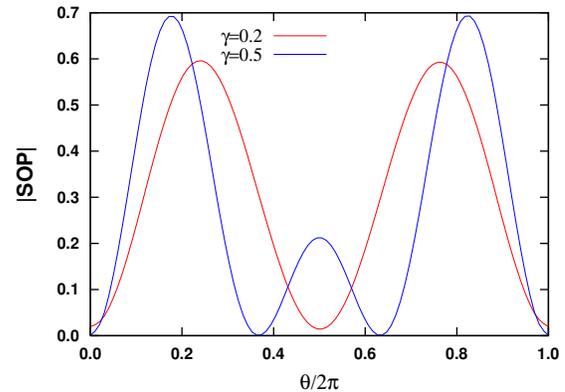}
\caption
{(Color online)
string order parameter $O^\textrm{even}_\textrm{str}(i=22,j=80,L=100,\theta)$ computed below and
above the critical point $\gamma_c=0.42$ of a $S=3/2$ alternating dimerized chain. The number of zeros
 of this parameter determines the nature of the valence
bond solid at each phase.
}
\label{SOP_theta1}
\end{figure}

Notice that, including the edges, both definitions involve an odd number of spins between sites
 $2i$ and $2j$ or $2i+1$ and $2j+1$ and an even number of antisymmetric operators under
spin flip. This condition is required to obtain a value of the mean value different from zero.

It has been commented in the previous sections that the $S=3/2$ alternating dimerized chain has three critical points
at $\gamma_c=0$ and $\gamma_c=\pm0.42$ in the interval $\gamma\in[-1,1]$. The phase diagram
is symmetric respect $\gamma=0$ and therefore we can constrain our study to the region 
$0\le\gamma$.

As regards the interval $0<\gamma<\gamma_c$ the ground state is known to be a $(2,1)$-VBS while 
for $\gamma_c<\gamma\le 1$ it is a $(3,0)$-VBS. The first issue we have to check is wether or not
the parameters defined above can make explicit this quantum phase transition. 
In fig. \ref{S32_SOP} (up) we have plotted both parameters 
$O^\textrm{odd}_\textrm{str}(\pi,L)$ and
$O^\textrm{even}_\textrm{str}(\pi,L)$ in the range $0\le\gamma\le1$. This graph shows that both
operators are finite at one side of the critical point while thay vanish at the other.
We can conclude then that they act as proper order parameters in the phase transition. However a
 clear major difference can be noticed from this figure since each one of these 
order parameters in fact vasnishes on differents sides of the critical point.

\begin{figure}
\includegraphics[scale=1.0]{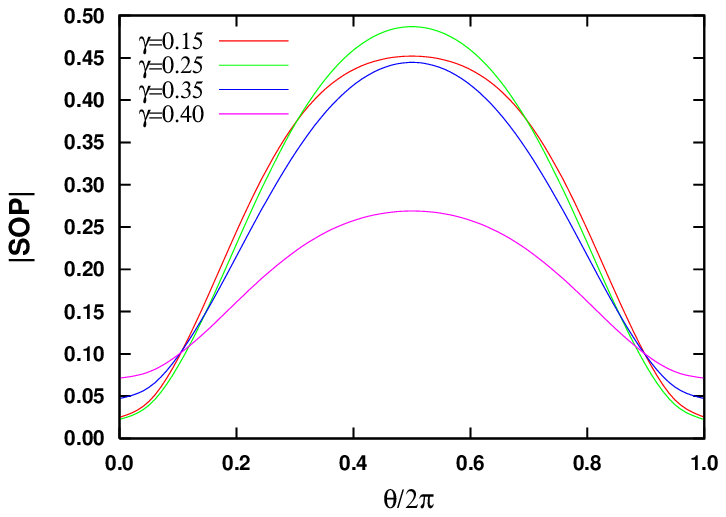}
\includegraphics[scale=1.0]{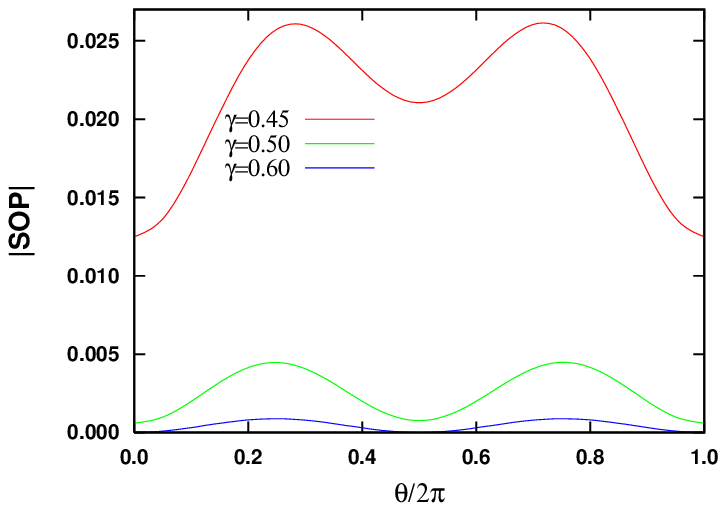}
\caption
{(Color online)
string order parameter $O^\textrm{odd}_\textrm{str}(i=21,j=79,L=100,\theta)$ computed below
(\textit{Up}) and
above (\textit{Down}) the critical point $\gamma_c=0.42$ of a $S=3/2$ alternating dimerized chain. In this case
 the string order parameter in the $\theta$ does not clearly determine the valence bond
 phases of the chain.
Notice also the appreciable change of scale from one phase to the other.
}
\label{SOP_theta2}
\end{figure}

On the other hand, the behaviour of the string order parameters at the critical point 
$\gamma=0$ requires further
 insight. We know that critical points correspond to critical ground states where the 
hidden order measured by string
order parameters must vanish. In fig. \ref{S32_SOP} (up) it is not however clear that the value
of the string order parameters decays to zero at this critical point. We have addressed
deeper this issue in fig. \ref{S32_SOP} (down). This graph shows a finite
 size scaling analysis of the
 string order parameter $O^\textrm{even}_\textrm{str}$ at the critical point $\gamma=0$.
 It can be observed that the scaling and
decay of the parameter is more influenced by the distance $i-j$ than the total length of
the chain. The rise of the string order parameter for large distances $i-j$ close
to the total lenght $L$ is expected due to finite size effects. For values far enough of
 the edges the tendency of the string order parameter is however to vanish as expected
 increasing 
the distance $i-j$. 
 
A regards the shape of these parameters for arbitrary $\theta$,
 fig. \ref{SOP_theta1}
shows $O^\textrm{odd}_\textrm{str}(\theta,L)$ for two different values of $\gamma$ 
below and above the critical point.From the knowledge that we have of the phases of
the chain we can see that this order parameter
behaves as pointed out in \cite{oshikawa92} and the number of zeros identifies the
valence bond phase ocurring. In effect, for values $\gamma<\gamma_c$ the number of vanishing
values in $0\le \theta<2\pi$ is two and coincides with a $(2,1)$-VBS while for
$\gamma >\gamma_c$ there are three null values corresponding to the $(3,0)$-VBS phase.

On the other hand $O^\textrm{even}_\textrm{str}(\theta,L)$ is plotted for
various values of $\gamma$ in fig. \ref{SOP_theta2}.
Two features can be remarked of this parameter: it also shows a qualitative change in its shape as we move from one phase
to the other. In this case however the relation of the number of zeros with the nature of the 
phase is not clear. Besides, the characteristic scale of the parameter differs significantly
 in both phases.


\begin{thebibliography}{99}
\bibitem{comment1}
With less precision since computations in fig. \ref{SOPvsGamma_aferro}
were performed in shorter ladders.
\end{thebibliography}
\end{document}